\documentclass[%
 reprint,
superscriptaddress,
showpacs,
nofootinbib,
 amsmath,amssymb,aps,
prd,
floatfix,
]{revtex4-1}

\usepackage{lipsum} 

\usepackage{bm}
\usepackage{comment}
\usepackage{floatrow} 
\pdfoutput=1
\usepackage{amsmath,amsfonts,amsthm}
\usepackage{esdiff}  
\usepackage{booktabs}  
\usepackage{url}  
\usepackage{hyperref}  
\usepackage{relsize}
\usepackage{cleveref}  
	\crefname{equation}{equation}{equations}
	\crefname{figure}{figure}{figures}	
	\crefname{table}{table}{tables}
\usepackage[caption=false]{subfig}

\usepackage[normalem]{ulem}

\usepackage[usenames,dvipsnames,svgnames,table]{xcolor}

\usepackage[export]{adjustbox}
\usepackage{graphicx}

\usepackage[sc]{mathpazo} 
\usepackage[T1]{fontenc} 
\usepackage{microtype} 

\usepackage{booktabs} 
\usepackage{float} 
\usepackage{hyperref} 

\usepackage{lettrine} 
\usepackage{paralist} 

\usepackage{titlesec} 
\renewcommand\thesection{\Roman{section}} 
\renewcommand\thesubsection{\Alph{subsection}} 
\titleformat{\section}[block]{\large\scshape\centering\bfseries}{\thesection.}{1em}{} 

\titleformat{\subsection}[block]{\scshape\centering}{\thesubsection.}{1em}{} 

\DeclareCaptionFormat{myformat}{#1#2#3\hrulefill}
\usepackage{float}
\floatstyle{plaintop}
\restylefloat{table}

\begin{document}

\title{Inference solves a boundary-value collision problem, with relevance to neutrino flavor transformation} 

\author{Eve Armstrong}
\email{evearmstrong.physics@gmail.com}
\affiliation{Department of Physics, New York Institute of Technology, New York, NY 10023, USA}
\affiliation{Department of Astrophysics, American Museum of Natural History, New York, NY 10024, USA}

\date{\today}

\begin{abstract}
Understanding neutrino flavor transformation in dense environments such as core-collapse supernovae (CCSN) is critical for inferring nucleosynthesis and interpreting a detected neutrino signal.  The role of direction-changing collisions in shaping the neutrino flavor field in these environments is important and poorly understood; it has not been treated self-consistently.  There has been progress, via numerical integration, to include the effects of collisions in the dynamics of the neutrino flavor field.  While this has led to important insights, integration is limited by its requirement that full initial conditions must be assumed known.  On the contrary, feedback from collisions to the neutrino field is a boundary value problem.  Numerical integration techniques are poorly equipped to handle that formulation.  This paper demonstrates that an inference formulation of the problem can solve a simple collision-only model representing a CCSN core -- without full knowledge of initial conditions.  Rather, the procedure solves a two-point boundary value problem with partial information at the bounds.  The model is sufficiently simple that physical reasoning may be used as a confidence check on the inference-based solution, and the procedure recovers the expected model dynamics.  This result demonstrates that inference can solve a problem that is artificially hidden from integration techniques -- a problem that is an important feature of flavor evolution in dense environments.  Thus, it is worthwhile to explore means of augmenting the existing powerful integration tools with inference-based approaches. 
\end{abstract}

\maketitle 

\section{Introduction} \label{sec:intro} 

The physics of flavor evolution in the neutrino field in high-density environments such as core collapse supernovae (CCSN) can significantly affect the transport of energy, entropy, and lepton number, with implications regarding nucleosynthesis, the mechanism of explosion, and the mass assembly histories of galaxies~\cite{mirizzi2016supernova,duan2010collective,chakraborty2016collective}.  Due to the fierce nonlinearity of various features of this problem, we are far from a complete understanding.   One important and poorly studied problem is the role of direction-changing collisions in shaping the neutrino density profile in these environments.

It is well established that coherent forward scattering of neutrinos with matter can lead to resonant effects~\cite{mikheev2007neutrino,mikheev1985resonance,wolfenstein1978neutrino}, and that forward scattering of neutrinos with each other also affects flavor evolution~\cite{fuller1987resonant,notzold1988neutrono,pantaleone1992neutrino,pantaleone1992dirac}.  These potentials, which govern neutrino flavor transformation, themselves depend on the flavor states of the neutrinos. This nonlinearity has led to the development of highly sophisticated numerical approaches, including the multi-angle Neutrino BULB code~\cite{duan2006simulation,duan2008simulating} and the IsotropicSQA code~\cite{richers2019neutrino}. 

The ability of direction-changing scattering, in addition, to significantly impact flavor evolution in the SN envelope was first shown by Ref~\cite{cherry2012neutrino}.  There have since been efforts to understand this "halo effect" on flavor transformation~\cite{cherry2013halo,cirigliano2018collective,zaizen2020neutrino,cherry2020time}, although a self-consistent solution has not been obtained.  More recent studies have shown that fast pairwise conversions can happen due to a crossing between the angular distributions of neutrinos and anti-neutrinos (See Ref~\cite{tamborra2020new} and citations therein) -- a phenomenon that is sensitive to the exact shapes of the angular distributions.  

Most recently, direction-changing collisions (hereafter "collisions") have been shown to enhance  fast flavor conversion in high-density environments~\cite{shalgar2021change,johns2021collisional,sasaki2021dynamics}.  This result is counter to the expectation that collisions should dampen flavor evolution via dynamical decoherence~\cite{martin2021fast}.  Moreover, even in small-scale CCSN models, accounting for collisions has yielded significant and counter-intuitive changes in the angular distributions of the neutrino density profiles.  For a more comprehensive summary of the role of collisions in flavor evolution, see Ref~\cite{shalgar2021change} and citations therein.



To date, efforts to include collisions in the neutrino quantum kinetic equations have used numerical integration to solve an initial value problem (e.g. Refs~\cite{cirigliano2018collective,shalgar2019occurrence,shalgar2021change}).  It is not necessarily desirable to place assumptions on initial conditions.  On the contrary, when feedback between collisions and flavor evolution is included in the calculation, the problem becomes a two-point boundary-value problem with no guarantee of a unique solution~\cite{shalgar2021change}. 

With that motivation, in this paper we solve a simple collision model, not as an initial-value problem, but rather as a two-point boundary value problem with partial information at the boundaries.  We do this via an inference -- or "inverse" -- formulation~\cite{tarantola2005inverse} of the problem.  Inference is a means to optimize a model with data, to predict model evolution at locations outside of those where the data have been provided.  The inference formulation does not require that initial conditions be known.  Rather, it requires that \textit{some} condition(s) (or constraint(s)) be known at \textit{at one or more} locations that parameterize the model dynamics.  These conditions may be placed at the endpoints of model evolution, or more generally at any point(s) along the model trajectory.  Further, the formulation of inference employed in this paper offers a mean to determine whether a unique solution exists, and -- if it does not -- the degree of degeneracy present.

In this paper, we use a formulation of inference -- statistical data assimilation (SDA) -- that is built to handle the case of extremely sparse data.  SDA was invented for numerical weather prediction~\cite{kimura2002numerical,kalnay2003atmospheric,evensen2009data,betts2010practical,whartenby2013number,an2017estimating}  and has gained considerable traction in neurobiology~\cite{schiff2009kalman,toth2011dynamical,kostuk2012dynamical,hamilton2013real,meliza2014estimating,nogaret2016automatic,armstrong2020statistical}.  Within astrophysics, inference has been used mainly for pattern recognition (e.g. ~\cite{Djorgovski2006}), while its utility for model completion is gaining notice in the exoplanet community~\cite{madhusudhan2018atmospheric} and solar physics~\cite{Kitiashvili2008}.  Recently, SDA has been applied to small-scale models of neutrino flavor evolution~\cite{armstrong2017optimization,armstrong2020inference,rrapaj2021inference}.  In those papers, the model contained coherent forward scattering only, and thus numerical integration could be taken as a confidence check on SDA solutions.

By contrast to Refs~\cite{armstrong2017optimization,armstrong2020inference,rrapaj2021inference}, this work tackles a boundary-value problem formulation, which numerical integration cannot access.  For this reason, we shall examine a simple model whose solution can be intuited via physical reasoning.  The model represents the high-density region of a supernova core, wherein neutrinos are effectively trapped and distributed isotropically.  In this context, we consider \textit{collisions alone}: absent neutrino-matter and neutrino-neutrino coupling potentials.  The model consists of four neutrino beams: two outgoing and two incoming, and four angular bins.  Absent flavor evolution, we will seek to examine the angular distributions of the neutrino number densities as functions of radius, due to collisions alone.  

Given partial information at the boundaries, the SDA procedure is able to recover the density profiles that follow from a simple physical argument, thereby illuminating physics that is artificially hidden from numerical integration techniques.  Further, multiple independent trials converge to one solution.

This paper proceeds as follows.  Section~\ref{sec:model} describes the model and imposed boundary conditions, and presents a simple physical argument regarding the expected model evolution.  Section~\ref{sec:method} describes the inference methodology and the simulated experiments performed.  Section~\ref{sec:result} shows how the SDA procedure predicts the expected result.  Finally, Section~\ref{sec:discussion} comments on the implications for tackling the full quantum kinetic equations that includes nonlinear feedback between collisions and flavor evolution, and advocates for exploring means to augment existing integration methodologies with inference-based techniques.

\section{Stationary box model of a supernova core} \label{sec:model}

The goal in crafting the model is twofold: 1) to create an opportunity to observe the effect on the neutrino angular distributions due to collisions alone, and 2) to retain a simplicity such that the solution can be intuited via physical reasoning, so as to provide a confidence check on the SDA solution.  To this end, we employ a two-dimensional flat "box" model and four neutrino beams: two incoming and two outgoing.

\begin{figure}[H]  
  \includegraphics[width=0.8\textwidth]{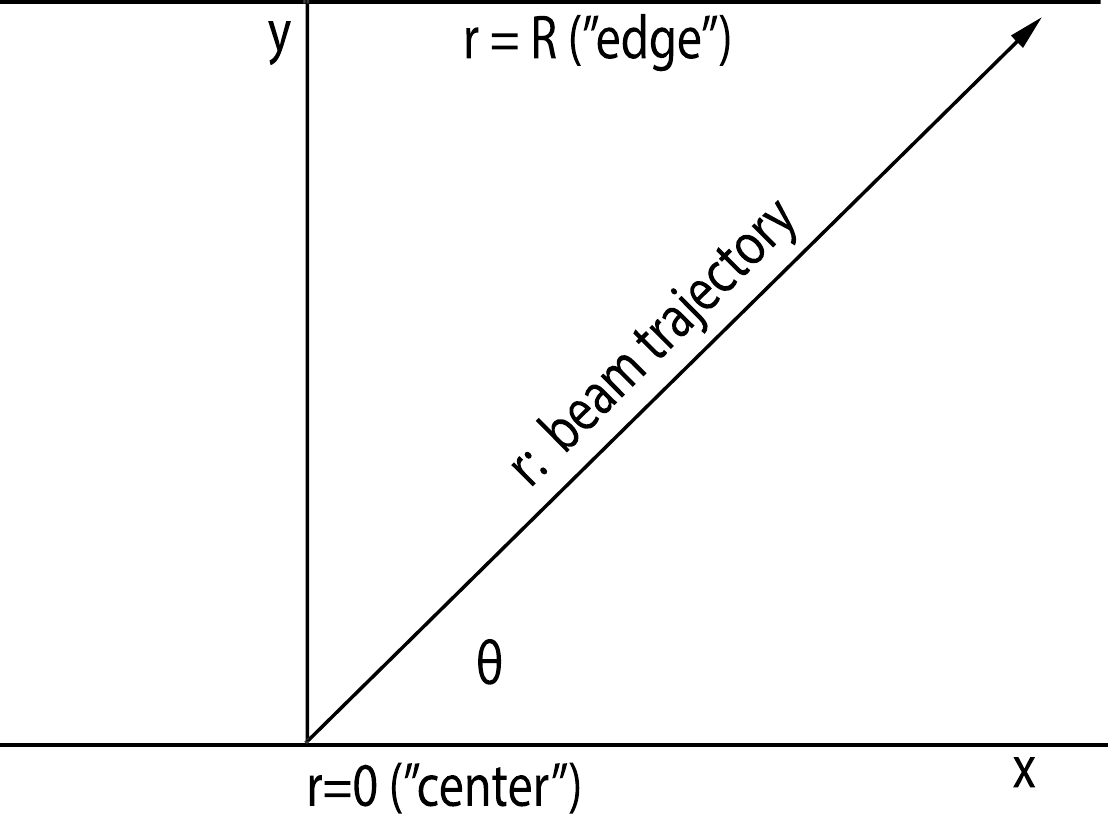}\\
  \caption{\textbf{Two-dimensional "box" model of the supernova core.}  The core center is the x-axis, at $r=0$.  The positive $\hat{y}$ direction is outward toward the edge.  Two beams are directed outward, and two inward, each along some angle $\theta$.}
\label{fig1}
\end{figure}

The model schematic of the supernova core is shown in Figure~\ref{fig1}.  Neutrinos radiate from the center (the x-axis at $y=0$) out to some final radius $R$ (at $y=y_{max}$).  To examine the effects of collisions alone, we omit flavor evolution and adopt a single-flavor model, with no anti-neutrinos.  There is a sink at $r=R$ where the outgoing beams may escape.  There is also a source and sink permitted at $y=0$, as particles are permitted to travel in the negative $\hat{y}$ direction.  

Neutrino flavor evolution can be expressed in terms of a density matrix $\rho$ (and $\overline{\rho}$ for anti-neutrinos).  For each neutrino momentum mode $\vec{p}$, we can write the following quantum-kinetic equation~\cite{volpe2015neutrino,vlasenko2014neutrino}:
\begin{align}
\begin{split}
  \left(\frac{\partial}{\partial t} + \vec{v} \cdot \nabla \right) \rho(\vec{x},\vec{p},t) = -i[H(\vec{x},\vec{p},t),\rho(\vec{x},\vec{p},t)]\\ + C\{\rho(\vec{x},\vec{p},t),\overline{\rho}(\vec{x},\vec{p},t)\},
\label{eq:qke}
\end{split}
\end{align}
\noindent where $\vec{x}$ and $t$ are spatial and temporal coordinates, respectively.  On the right side, the Hamiltonian term $H$ describes neutrino mixing in vacuum, neutrino interactions with the matter background, and neutrino-neutrino self interactions; it contains coherent scattering.  The second term $C$ is the "collision term," which accounts for direction-changing scattering\footnote{Note that the separation of coherent from direction-changing scattering is an artificial one; see Section~\ref{sec:discussion}.}.

In this paper, we neglect the (coherent) Hamiltonian term and focus on the collision term $C$.  We shall assume that a steady-state can be defined on a timescale shorter than the leakage timescale, and so we eliminate time from Equation~\ref{eq:qke}.  That is, we will solve Equation~\ref{eq:qke} depending on scattering angle and not on momentum.  In this context, the velocity in the second operator on the left side shall be considered a spatial -- rather than a temporal -- velocity.  Then our equation of motion simplifies to:
\begin{equation}
\textrm{cos} \theta_i \diff{\rho_i(r)}{r} = \sum_j \left(-C^{\textrm{loss}}\rho_i(r) + C^{\textrm{gain}}\rho_j(r)\right)\frac{\Delta \textrm{cos} \theta_j}{2}.  
 \label{eq:collisions}
\end{equation}
\noindent Here, subscripts ($i,j$) denote indices of angular bins, $\Delta \textrm{cos}\theta_j$ is the width of the jth angular bin, and the cosine term on the left side accounts for the dependence of path length on zenith angle.  $C^{\textrm{loss}}$ and $C^{\textrm{gain}}$ are the coefficients for loss and gain for each bin; they represent the strength of interaction between neutrinos and nucleons\footnote{Physically, the loss and gain coeffients represent the product of the number of scatterers and the cross-section for each particular interaction channel averaged over some distribution of neutrino energy~\cite{shalgar2019occurrence}.}.  Thus for our box model, there are four Equations~\ref{eq:collisions}, each governing the evolution of the neutrino density in one of four angular bins.

To create a problem with a physically intuitable solution, we make further simplifications.  First, the coefficients $C^{\textrm{loss}}$ and $C^{\textrm{gain}}$ are constant numbers throughout the medium.  That is, the matter density is constant within the box and drops discontinuously to zero at $r=R$.  Next we set $C^{\textrm{loss}}$ equal to $C^{\textrm{gain}}$, to conserve particle number at each interaction\footnote{Note that setting $C^{\textrm{loss}}$ equal to $C^{\textrm{gain}}$ renders Equation~\ref{eq:collisions} trivial, unless there are slight asymmetries in the values of the $\textrm{cos}\theta$ terms on the left side of the equations.  For this reason, we introduce slight asymmetries between the fluxes of outgoing versus incoming beams.  The $\textrm{cos}\theta$ values are 0.8 and 0.9 for the outgoing beams, and -0.6 and -0.7 for those incoming.}.

Finally, we choose a value of $C$ to approximate nuclear density.  Near nuclear density, neutrinos are effectively trapped, and so we expect an isotropic distribution of scatterers: at each collision, a neutrino has equal probability of scattering inward as outward.  To choose $C$ appropriately, we perform a dimensional analysis of Eq~\ref{eq:collisions}.  The left side is a derivative, with units of 1/length.  The right side also has units of 1/length, as the only dimensionful quantities on the right side are the coefficients $C$, with units of inverse length.  Physically, as $C$ represents the strength of the interaction between neutrinos and matter, $C$ can be considered the inverse of the neutrino mean free path, with units of $m^{-1}$.  For example, taking a mean free path on the order of meters to represent nuclear density, we choose $C$ to be 0.2 $m^{-1}$.

\subsection{The boundary conditions}

We place on this model partially-known boundary conditions.  First, the values of the outgoing beams at the core center ($r=0$) are known.  At that location, we give them both a (dimensionless) number density of 0.5.  Second, the values of incoming beams at the core edge ($r=R$) are zero.  That is, by definition, at $r=R$, there occurs no further back-scattering.  This is the sole information furnished to the inference procedure.  The procedure is tasked with taking this information together with the model equations of motion (Eq~\ref{eq:collisions}), to predict the complete density profile at all other spatial locations for all four angular bins --  including the values at $r=0$ of the incoming beams and the values at $r=R$ of the outgoing beams.

\subsection{Expected density profile, to be recovered by the SDA procedure}

Given a value of $C$ within our box that corresponds to nuclear density, and the fact that a system represented as ordinary differential equations has no memory, we expect to observe isotropy for the vast majority of the trajectories of all four beams.  That is, at each interaction the particles possess no preferred direction.  Thus, given the condition that the outgoing beams begin their trajectory at $r=0$ with a value of 0.5, we expect the densities of all beams to remain approximately 0.5 throughout the majority of their their trajectories.  Then, as we approach $r=R$, we should see a precipitous drop.  Near $R$ the densities of the incoming beams should drop to zero, in accordance with the imposed boundary condition there, and the inference procedure should \textit{predict} that the densities of the outgoing beams drop to some number below 0.5 -- that number to scale inversely with the value of $C$.  Stated physically: as $C$ weakens, more particles should escape.

Note again that the model possesses a leak at $r=R$ (and at $r=0$, as scatterers may pass beyond zero -- in the negative $\hat{y}$ of Figure~\ref{fig1}.  We do not expect that a steady-state solution exists for longer than a duration corresponding to the leakage timescale.  Rather, we expect that different choices for the coefficient $C$ will yield different solutions, each representing one freeze-frame of a steady state solution.  Specifically, as $C$ decreases, the mean free path increases -- and equivalently the total duration increases, as increasingly more neutrinos have had time to escape the core. 

\section{Inference methodology} \label{sec:method}

Statistical data assimilation (SDA) is an inference procedure in which a dynamical system is assumed to underlie any measured quantities. This model $\bm{F}$ can be written as a set of \textit{D} ordinary differential equations that evolve in some parameterization $r$ as:
\begin{align*}
  \diff{x_a(r)}{r} &= F_a(\bm{x}(r),\bm{p}(r)); \hspace{1em} a =1,2,\ldots,D,
\end{align*}
\noindent
where the components $x_a$ of the vector \textbf{x} are the model state variables.  Any unknown parameters to be estimated are contained in $\bm{p}$, and may themselves be variable.  In this paper, model 
$\bm{F}$ is the set of four equations (Eq~\ref{eq:collisions}) governing the neutrino density profile in each of four angular bins.  

A subset $L$ of the $D$ state variables is associated with measured quantities.  In this paper, the "measurements" are the four boundary conditions placed on the outgoing beams at $r=0$ and incoming beams at $r=R$.  One seeks to estimate the evolution of all state variables that is consistent with the measurements provided, to predict model evolution at parametrized locations where measurements are not present.  

We formulate the SDA procedure as an optimization wherein a cost function is extremized.  An optimization formulation of inference does not require knowledge of initial conditions.  Rather, it requires that \textit{some} constraints be placed on the model at \textit{some} location(s) on the coordinate axis that parameterizes the model equations of motion.  These conditions may exist at the endpoints of the problem, as is the case in this paper.  Importantly, however, the conditions need not be at the bounds.  This flexibility is the key advantage of inference, compared to integration, that we aim to demonstrate.

The cost function is written in two terms\footnote{Additional terms representing equality constraints may be added to the cost function, depending on the aim of a particular optimization procedure.}.  One term represents "measurement error":  the difference between state prediction and any measurements made.  The second term represents "model error": the difference between state prediction and adherence to the model dynamics.  It will be shown in Section~\ref{sec:result} that treating the model error as finite offers a means to determine whether a particular solution is consistent with both measurements and model dynamics, as well as a means to assess uniqueness.  

We search the surface of the cost function via the variational method.  The procedure in its entirety - that is: a variational approach to minimization coupled with an annealing method to identify a lowest minimum of the cost function (which will be described below in this Section) - is referred to as variational annealing (VA).  The procedure searches a $(D \,(N+1)+ p)$-dimensional state space, where \textit{N} is the number of discretized steps.  One seeks the path $\bm{X}^0 = {\bm{x}(0),...,\bm{x}(N),\bm{p}(0),...\bm{p}(N)}$ in state space on which the cost function attains a minimum value.  We refer to the cost function $A_0$ as the action, because it can be derived from the concept of a classical action of a particle on a path in a state space~\cite{abarbanel2013predicting}.  Ref~\cite{armstrong2020inference} demonstrated that the action formulation offers a simple litmus test for identifying correct solutions: namely, they are solutions that correspond to the path of least action.

The cost function $A_0$ used in this paper is written as:
\begin{widetext}
\begin{equation} \label{eq:action}
\begin{split}
A_0 =& R_f A_{model} + R_m A_{meas}\\
A_{model}=&\frac{1}{{N}D}	\mathlarger{\sum}_{n \in \{\text{odd}\}}^{N-2} \mathlarger{\sum}_{a=1}^D \left[ \left\{x_a(n+2) - x_a(n) - \frac{\delta r}{6} [F_a(\bm{x}(n), \bm{p}) + 4F_a(\bm{x}(n+1),\bm{p}) + F_a(\bm{x}(n+2),\bm{p})]\right\}^2  \right. \\
  & \hspace{100pt} + \left.\left\{ x_a(n+1) - \frac12 \left(x_a(n)+x_a(n+2)\right) - \frac{\delta r}{8} [F_a(\bm{x}(n),\bm{p}) - F_a(\bm{x}(n+2),\bm{p})]\right\}^2 \right] \\
  A_{meas}=& \frac{1}{N_\text{meas}} \mathlarger{\sum}_j \mathlarger{\sum}_{l=1}^L (y_l(j) - x_l(j))^2.
\end{split}
\end{equation}
\end{widetext}

The second (simpler) squared term of Equation~\ref{eq:action} governs the transfer of information from measurements $y_l$ to model states $x_l$\footnote{This term derives from the mutual information of probability theory~\cite{abarbanel2013predicting}.}.  Here, the summation on \textit{j} runs over all discretized locations $J$ at which measurements are made, which may be some subset of all discretized steps of the model.  The summation on \textit{l} is taken over all \textit{L} measured quantities.  In the simulations of this paper, the measured quantities are the number densities in the angular bins, the outgoing beams sampled at  radius of $r=0$ (at $j=1$ only), and the incoming beams sampled only at $r=R$ (at the value of $j$ that denotes the final radial location $R$).

The first squared term of Equation~\ref{eq:action} incorporates the model evolution of all \textit{D} state variables $x_a$.  Here, the outer sum on \textit{n} is taken over all discretized timepoints of the model equations of motion.  The sum on \textit{a} is taken over all \textit{D} state variables at all discretized locations.\footnote{This term can be derived via consideration of Markov-chain transition probabilities~\cite{abarbanel2013predicting}.}  In our model, these measured quantities are all four beams, or $L = D = 4$.  The first and second bracketed terms represent error in the first and second derivative of the model, respectively.  For further details on the action formulation, see \textit{Appendix A} of Ref~\cite{armstrong2017optimization}.  

Importantly, note that the SDA procedure is tasked with inferring from the sparse boundary conditions the complete model evolution at \textit{all} locations on $r$.  

Finally, if the model is nonlinear, the cost function surface will be non-convex: multiple minima may exist.  For this reason, we employ an annealing procedure to identify the global minimum of the problem, and to ascertain whether there exists a single global minimum -- that is, a unique solution.  Details of this annealing procedure will be presented in Section~\ref{sec:result}.  

\subsection{Details of the simulated experiments}

We performed the simulated experiment for three distinct values of the coefficient $C^{\textrm{loss}}=C^{\textrm{gain}}$: 2.0, 0.2, and 0.02 $m^{-1}$, where the choice of 0.2 $m^{-1}$ represents nuclear density (a mean free path of 5 meters).  At each discretized model location $r$, the procedure was permitted to search the full dynamical range of each state variable of [0:1].  For each of those three experiments, we took the size of the core to be $R=50 km$, and examined the robustness of the solution across three distinct choices for step size $dr$: 1 m, 10 m, and 100 m\footnote{Within the dimensionless mathematical framework of the SDA procedure, those choices of 1, 10, and 100 $m^{-1}$ corresponded to: a step size of 0.01 with 10101 steps, 0.1 with 1011 steps, and 1.0 with 101 steps, respectively -- keeping to a total size of 101.}.  

To examine the uniqueness of solutions, for each of the (nine) experiments described, 20 independent paths were searched, each beginning at a randomly-generated set of initial guesses for the four state variables at each parameterized location $r$.

To perform the optimization, we used the open-source Interior Point Optimizer (Ipopt)~\cite{wachter2009short}.  Ipopt uses a Simpson’s rule method of finite differences to discretize the state space, a Newton’s method to search, and a barrier method to impose user-defined bounds that are placed upon the searches.  The discretization of the state space, the calculations of the model Jacobean and Hessian matrices, and the annealing procedure were performed via an interface with Ipopt that was written in C and Python~\cite{minAone}.  All simulations were run on a 720-core, 1440-GB, 64-bit CPU cluster.

\section{Result} \label{sec:result} 

\subsection{General findings}

Results are fourfold. 
\begin{itemize}
  \item As expected for an isotropic distribution of scatterers, for a large value of the coefficient $C^{\textrm{loss}} = C^{\textrm{gain}}$, the number densities of the four angular bins remain constant at 0.5 for the majority of the trajectory, with a precipitous drop near the bound at $r=R$.  As the value of $C$ decreases, the rate of depletion of the number densities as a function of $r$ slows, and the predicted values for the outgoing beams at $r=R$ show that the densities in those bins have been depleted to a lesser degree, as more particles escape the core.  In the limit of no collisions ($C=0$), the solution to Equation~\ref{eq:collisions} is trivial and all particles escape.   
  \item The results are invariant, to one part in $10^6$, across choices of discretized step size that span three orders of magnitude.  
  \item When the procedure receives one additional constraint such that the rates of depletion of the bin populations near $R$ are captured, and meanwhile the value of $C$ is withheld, the procedure is able to correctly \textit{infer} the value of $C$ that corresponds to those rates of depletion. 
  \item For all simulations performed, the plot of the cost function over the course of an annealing procedure (to be described below) shows that i) all results converge; that is: solutions are consistent with both boundary conditions and model dynamics, and ii) each solution is unique; that is: all paths searched converge to one solution. 
\end{itemize}

\subsection{Effect of collision strength $C$}

\begin{figure*}[htb] 
  \includegraphics[width=17cm]{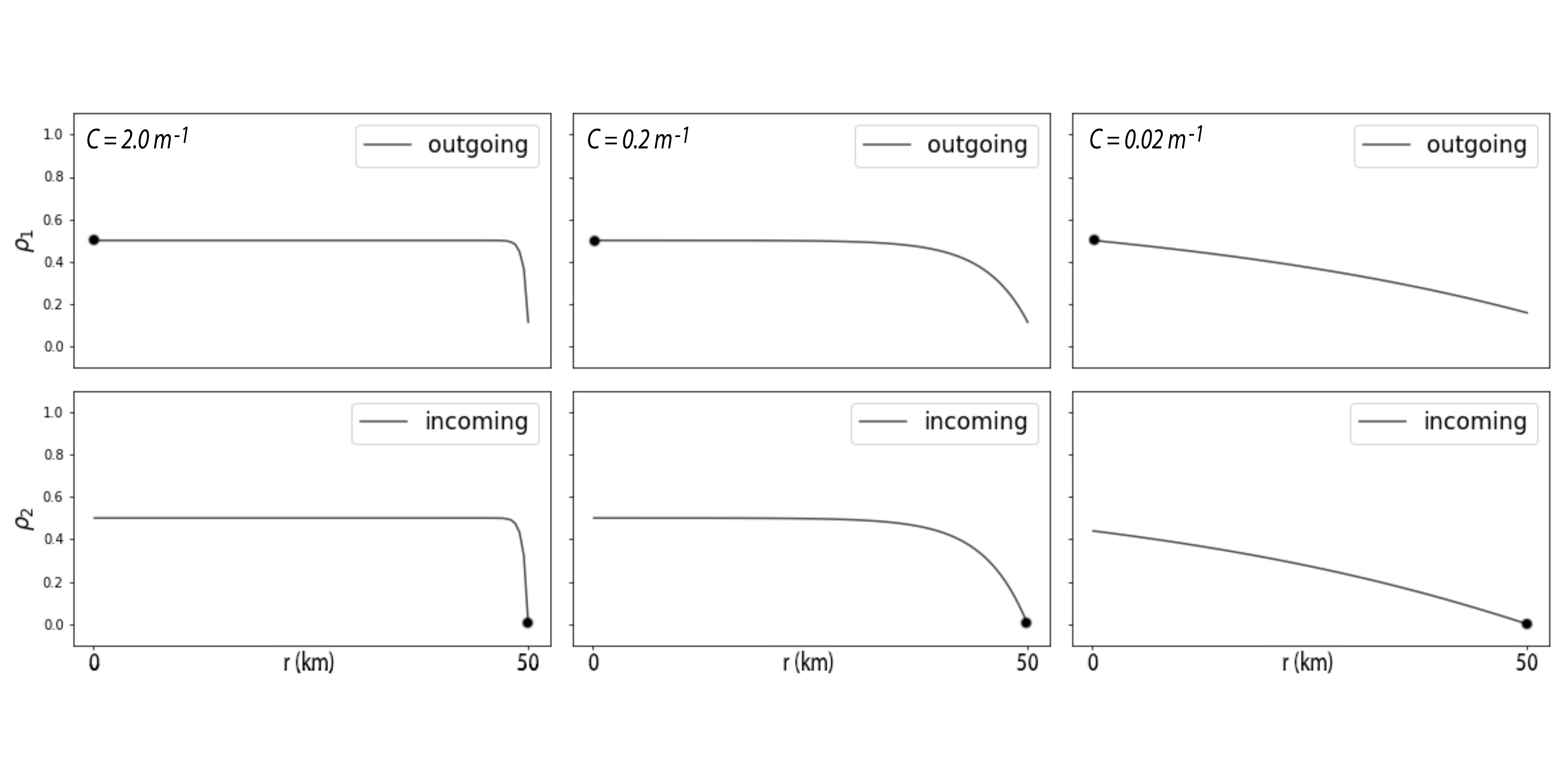}\\
  \caption{\textbf{SDA result for the density profiles as functions of radius $r$ of one outgoing (top) and one incoming beam (bottom), for three values of coefficients $C$.}  For left, middle, and right columns: $C =$ 2.0, 0.2, and 0.02 $m^{-1}$, respectively.  Circles denote locations of "measurements", or boundary conditions, provided to the procedure.  The remaining trajectories are the SDA prediction.  See the text for an interpretation in light of the expectation set forth in Section~\ref{sec:model}.}
 \label{fig2}
\end{figure*}

We examined the solution for three distinct values of collision strength $C^{\textrm{loss}} = C^{\textrm{gain}}$, to determine whether the optimization would recover our expectation that the rate of depletion of the number densities in the angular bins should slow as $C$ weakens.   Figure~\ref{fig2} shows the result.  The left, middle, and right columns correspond to values of $C$ of 2.0, C=0.2, and 0.02 $m^{-1}$, respectively.  The top and bottom rows show the density profile as a function of $r$ for one outgoing beam $\rho_1$ and one incoming beam $\rho_2$, respectively.  (The trajectories of the other two beams in the four-beam model are identical; not shown.)  The circles denote the locations of the "measurements", or boundary conditions, that were given as information to the SDA procedure.  The remainder of the trajectories are the SDA prediction, given knowledge of the model dynamics.  

From left to right in Figure~\ref{fig2}, note three effects.  First, the number densities near $r=0$ appear roughly the same, which follows from the argument regarding isotropy presented in Section~\ref{sec:model}.  

Second, as $C$ weakens, the number densities fall off near $r=R$ increasingly gradually.  Specifically, the predicted value for the outgoing beam $\rho_1$ at $r=R$ is 0.114 (left), 0.115 (middle), and 0.158 (right).  This follows from the expectation that as the mean free path lengthens, more particles escape.  

Third, as $C$ weakens, fewer incoming particles are predicted to reach $r=0$.  Specifically, the value for the incoming beam $\rho_2$ at $r=0$ is 5.0 (left), 4.9 (middle), and 4.4 (right).  These numbers are summarized in Table~\ref{table1}. 

Recalling that the model contains a leak at $r=R$, we can consider the left, middle, and right panels of Figure~\ref{fig2} to be a succession of solutions in time, each approximating a steady state over a timescale shorter than the leakage timescale.  As $C$ weakens from left to right, the time increases.  In the limit where $C=0$, the solution to Equation~\ref{eq:collisions} is trivial, and all particles escape as time goes to infinity (not shown).

Finally, we make several notes regarding the robustness of the results.  The procedure imposes the boundary conditions ($\rho_1 = 0.5$ at $r=0$ and $\rho_2 = 0.0$ at $r=R$) to one part in $10^{7}$.  The results are invariant, to one part in $10^6$, across the three choices of discretized step size $dr$ corresponding to 1, 10, and 100 m\footnote{For the reader interested in mapping these choices to a realistic physical scenario: a choice of $C = 0.2  m^{-1}$, which corresponds to a mean free path of 5m, together with a step size $dr$ of 1 m, would adequately sample a particle's path.}.  The limit on that accuracy is likely due to a discretization error internal to Ipopt, the optimizer used in the procedure.   


\setlength{\tabcolsep}{5pt}
\begin{table}[H]
\small
\centering
\begin{tabular}{c | c c c } \toprule
 \textit{C} ($m^{-1}$) & 2.0 & 0.2 & 0.02\\\midrule \hline\hline
 \textit{$\rho_1$} at $r=R$ (outgoing) & 0.114 & 0.115 & 0.158\\
 \textit{$\rho_2$} at $r=0$ (incoming) & 0.5 & 0.49999 & 0.44\\\bottomrule
\end{tabular}
\caption{\textbf{Predicted values of outgoing and incoming beams at the endpoints, across values of $C$.}  As $C$ dilutes from left to right, \textit{top}: more outgoing neutrinos escape to $r=R$; \textit{bottom}: fewer incoming neutrinos return to $r=0$.}
\label{table1}
\end{table}

\subsection{Correct inference of collision strength $C$ as an unknown parameter}

Having determined that the SDA procedure can predict the rate of depletion of the angular bins given a particular strength of coefficients $C$, we sought to determine whether it could perform the inverse: to infer $C$ given a rate of depletion.

To this end, we repeated the experiment leaving $C$ as an unknown parameter to be estimated.  As before, we gave the SDA procedure the original four boundary conditions.  In addition, from the original solutions shown in Figure~\ref{fig2} we gave to the procedure the values of densities in each angular bin at five discretized locations prior to $R$, with the aim of capturing the rate of depletion of the bins.  For each of the three cases, the procedure correctly estimated the unknown value of $C$ to sixth-decimal precision.

\subsection{Convergence and uniqueness}

For a nonlinear model, the surface of the action, or cost function, will be non-convex.  The complete SDA procedure anneals in terms of the ratio of model and measurement error, $R_f$ and $R_m$, respectively\footnote{More generally, $R_m$ and $R_f$ are inverse covariance matrices for the measurement and model errors, respectively.  In this paper the measurements are taken to be mutually independent, rendering these matrices diagonal.}, to gradually freeze out a lowest-minimum of the Action~\cite{ye2015systematic}.  This iteration works as follows.

We define the coefficient of measurement error $R_m$ to be 1.0, and write the coefficient of model error $R_f$ as: $R_f = R_{f,0}\alpha^{\beta}$, where $R_{f,0} = 10^{-1}$, $\alpha = 1.5$, and $\beta$ is initialized at zero.  Parameter $\beta$ is the annealing parameter.  When $\beta = 0$, relatively free from model constraints the Action surface is smooth and there exists one minimum of the variational problem that is consistent with the measurements.  We estimate that minimum.  Then we increase the weight of the model term slightly, via an integer increment in $\beta$, and recalculate the action.  We do this recursively toward the deterministic limit of
\begin{figure}[H] 
  \includegraphics[width=8cm]{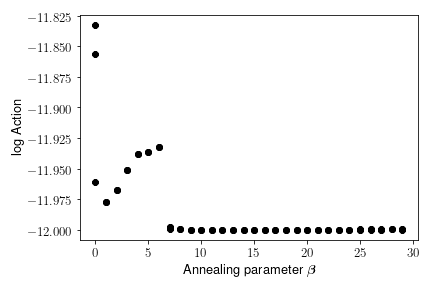}
  \caption{\textbf{Action, or cost function, as function of annealing parameter $\beta$, for a value of $C$ of 0.2}; representative of all simulations.  As $\beta$ passes 7, the action flatlines, indicating convergence to a solution that is consistent with both measurements (or boundary conditions) and model dynamics.} 
 \label{fig3}
\end{figure}
\noindent $R_f \gg R_m$.  The aim is to remain sufficiently near to the lowest minimum so as not to become trapped in a local minimum as the surface acquires the structure imposed by the model dynamics.

Ref~\cite{armstrong2020inference} demonstrated that, for a particular path searched, the plot of action as a function of annealing parameter $\beta$ indicates whether a particular solution is stable as the model dynamics are imposed increasingly rigidly.  Figure~\ref{fig3} shows this result for the case of $C=0.2$, representative of all experiments performed in this paper.  Note that at a value of $\beta = 7$, the action flatlines.  This behavior indicates that a solution has been found that is consistent with both imposed boundary conditions and with model dynamics.  For a detailed explanation, see Ref~\cite{armstrong2020inference}.

Finally, the 20 paths initialized for each experiment in this paper each converged to a single solution.  This was demonstrated by identical state variable evolution (e.g. Figure~\ref{fig2}) and identical action($\beta$) plots (e.g. Figure~\ref{fig3}) for each path sampled.  Of course, to ascertain whether a solution is truly unique, one must sample an infinite number of paths.  The specific number chosen for any particular model should depend on that model's complexity and dimensionality.

\section{Discussion} \label{sec:discussion}

We have considered a simple two-point boundary value collision model with partially-known boundary conditions, and challenged an inference procedure to solve it.  The model is sufficiently simple that a physical argument can intuit the expected result, thereby serving as a confidence check on the inference-based solution.  Specifically, an optimization formulation of SDA recovers the expected angular distributions of the density profiles of particles in a collision-only flat two-dimensional model, where the particles represent neutrino-nucleon interactions near nuclear density.  The behavior of the cost function over an iterative reweighting of the terms that impose the boundary conditions and the equations of motion, respectively, demonstrates that the SDA solution is consistent with both the boundary conditions and the model dynamics.  This was the case over a range of choices for the collision strength that spanned three orders of magnitude, and the results are robust to the step size used for discretization.  Multiple randomly-initiated searches of the state-and-parameter space converged to one solution.

This finding has important implications regarding the potential of inference to inform flavor transformation models governed by the full quantum kinetic equation (Equation~\ref{eq:qke}; that is, taking the Hamiltonian together with direction-changing collisions.  While the integeration-based approaches to the collision problem have led to valuable insights, considering the full feedback between collisions and flavor evolution, will require that the problem be recast as a boundary problem with no guarantee of a unique solution~\cite{shalgar2021change}.  An inference formulation avoids reliance on initial conditions, and rather seeks a solution that is compatible with constraints placed only at locations where we have high confidence in our understanding of the physics there.  In the simulations performed in this paper, those constraints were placed at the endpoints, but they need not be.  Further, within the optimization formulation, uniqueness can be investigated via the initialization of  multiple independent searches.  The ability of this SDA procedure to probe model degeneracy has been demonstrated in significantly more detail in Refs~\cite{armstrong2017optimization,armstrong2020inference}. 


It is relevant to note here another potential advantage of inference for augmenting numerical integration tools: efficiency.  Including feedback from collisions self-consistently in the quantum kinetic calculations dramatically increases the computational complexity, as both the collisions and the flavor evolution will simultaneously shape the neutrino angular distributions~\cite{shalgar2021change}.  The simple experiment presented in this paper did not offer the opportunity to showcase the high efficiency with which this SDA formulation performs state-and-parameter estimation; it has been demonstrated for a small-scale flavor evolution model in Ref~\cite{armstrong2020inference}.  


The question of \textit{how} to implement inference within the existing numerical integration framework has yet to be explored.  It is relevant here to note that the separation of coherent versus direction-changing scattering in Equation~\ref{eq:qke} -- the former into the Hamiltonian and the latter into the collision term $C$ -- is an artificial choice.  It might be worthwhile to consider possible advantages of rewriting the quantum kinetic formulation to treat coherent and direction-changing scattering \textit{collectively}.  Of course, this is a daunting suggestion, as the the formulation of Equation~\ref{eq:qke} forms the foundation of extremely large and powerful codes.  Nevertheless, it is worth asking to what extent that formulation limits our ability to access the full dynamical range of the flavor field in compact object environments.  

In parallel with exploring means to fold inference into the existing codes, it will be instructive to continue examining how inference performs on collision-only models of increasing complexity.  For example, in the model used in this paper, neutrino number density was conserved at each collision: collisions were completely elastic and did not absorb or emit neutrinos.  Further, the matter density was held constant, with a discontinuous fall to zero at the core's edge.  A more realistic matter profile will contain some radial dependence.  In addition, model behavior should be examined within a spherically-symmetric geometry.  Because adding complexity will obscure a simple expectation regarding the result, it will be instructive to identify alternative comparisons to inference-based solutions, such as random-walk formulations.

\section{ACKNOWLEDGEMENTS}

Thank you to Shashank Shalgar for invaluable conversations, without which this manuscript would not exist.  Thank you also to George Fuller and Amol Patwardhan for important comments.  E.~A. acknowledges the National Science Foundation (NSF grant 2139004), an Institutional Support for Research and Creativity grant from New York Institute of Technology, and Gretel Yeager.  

\bibliography{bib_collisions,bib_armstrong2020paper}

\end{document}